# Inverse melting and re-entrant transformations of the vortex lattice in amorphous Re$_6$Zr thin film

Rishabh Duhan[*], Subhamita Sengupta[*], John Jesudasan, Somak Basistha and Pratap Raychaudhuri[1]

*Tata Institute of Fundamental Research, Homi Bhabha Rd, Mumbai 400005, India.*

**Melting of a solid is one of the most ubiquitous phenomena observed in nature. Most solids, when heated, melt from a crystalline state to an isotropic liquid at a characteristic temperature. There are however situations where increase in temperature can induce a transition to a more ordered state[1]. Broadly termed as "inverse melting", experimental realisations of such situations are rare[2,3,4]. Here, we report such a phenomenon in the 2-dimensional vortex liquid that forms in a moderately pinned amorphous Re$_6$Zr (*a*-ReZr) thin film, from direct imaging of the vortex lattice using a scanning tunnelling microscope. At low temperature and magnetic fields, we find that the vortices form a "pinned liquid"[5], that is characterised by a low mobility of the vortices and vortex density that is spatially inhomogeneous. As the temperature or magnetic field is increased the vortices become more ordered, eventually forming a nearly perfectly ordered vortex lattice. Above this temperature/magnetic field, the ordered vortex lattice melts again into a vortex liquid. This re-entrant transformation from a liquid to solid-like state and then back to a liquid also leaves distinct signature in the magnetotransport properties of the superconductor.**

[1] pratap@tifr.res.in
[*]These authors contributed equally to the work.

In the mixed state of a clean Type II superconductor, the repulsive interactions between vortices are expected to arrange them in a hexagonal periodic lattice, the Abrikosov vortex lattice[6]. However, in real systems this tendency towards ordering competes against several other factors. Random pinning from structural defects that inevitably exist in a material tend to destroy the order in a periodic vortex solid. Thermal and quantum fluctuations can also melt the solid into a vortex liquid (VL). Consequently, except for very clean superconductors[7], the periodic vortex lattice is observed well below the upper critical field, $H_{c2}$, and critical temperature, $T_c$. As magnetic field or temperature is increased the periodic order is eventually destroyed, giving rise to a completely disordered state[8,9,10,11,12,13] (Fig. 1(a)). In conventional bulk superconductors where fluctuations are not strong enough to melt the vortex lattice, this disordered state is usually amorphous vortex glass[14,13,11]. Among bulk superconductors, VL has been unambiguously detected only in layered High-$T_c$ cuprates[15,16,17,18], at high operating temperatures. On the other hand, in thin films where the role of fluctuations is much stronger, VL states have been observed in many systems[19,20,5]. In this context, an interesting possibility is that of re-entrant melting. It has been proposed in numerous theoretical studies[21,22,23,24,25] that at very low fields, the soft vortex lattice could melt due to fluctuations giving rise to a VL. As the vortex lattice becomes more rigid with increase in field, this low-field VL transforms into a crystalline vortex solid, and then melt again into a VL at higher fields. More interestingly, many of these calculations suggest that there is a small region of phase diagram at low fields, where the liquid-solid-liquid re-entrant behaviour could also be observed as a function of increasing temperature (Fig. 1(b)). While there have been several attempts observe signatures of re-entrant melting from magnetic or transport measurements in single crystals of both conventional[26,27,28] and layered High-$T_c$ superconductors[29,30], these efforts have only been partially successful due to the formation of vortex glass states. In High-$T_c$ cuprates, pinning transforms the VL into a vortex glass below about $0.4T_c$. Since vortex glass states have large thermomagnetic irreversibility, experimental evidence of re-entrant melting is either confined to a small region at high temperatures[30,31] or employ indirect methods such as shaking the amorphous vortex glass with an a.c. magnetic field[29] to unravel the behaviour of the underlying VL state. In conventional low-$T_c$ superconductors, re-entrant order-disorder transformations of the vortex lattice have also been reported[14,26,27,28], but here both low-field and high-field disordered states are vortex glass instead of a VL. Till date, the re-entrant transformation of a vortex lattice from a vortex liquid to a vortex solid and back to a liquid both as a function of temperature and as a function of temperature has not been experimentally observed in any compound.

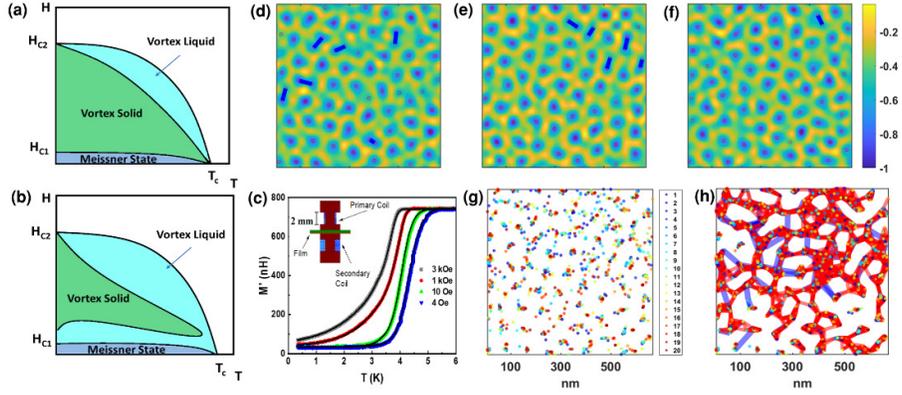

**Fig 1| Formation of inhomogeneous VL at low field. (a)** Conventional phase diagram of Type II superconductors. **(b)** Phase diagram including the low-field vortex liquid state. **(c)** Magnetic screening response measured from the real part of mutual inductance using the two-coil mutual inductance setup shown in the inset. The points correspond of measurements starting from a Zero field cooled (ZFC) state and the solid lines are the corresponding measurements starting from Field Cooled (FC) state. **(d-f)** Representative normalized conductance maps $G_N(r)$ from a series of 20 successive conductance maps acquired at 3 kOe, T = 460 mK; the three panels correspond to the 4$^{th}$, 10$^{th}$ and 16$^{th}$ images in this sequence. The local minima are shown by red circles. Blue lines join two adjacent shallow minima as described in the main text. **(g)** Positions of minima from 20 consecutive images and **(h)** global motion path of vortices in the inhomogeneous vortex liquid, at 3 kOe, 460 mK.

In very weakly pinned thin films, a vortex solid with well-established hexagonal order is observed in small region of the phase diagram at low fields and melts into a variety of VL states with isotropic, hexatic or smectic correlations at higher fields[20,19]. With increase in pinning the vortex solid phase further shrinks. When pinning is strong, an inhomogeneous VL forms over the entire *H-T* parameter space from very low magnetic field/temperatures all the way up to $T_c/H_{c2}$, as recently seen on few nanometers thick *a*-ReZr and *a*-MoGe films[5]. Therefore, an interesting question to ask is how the vortex state will evolve with temperature and magnetic field when the pinning is considerably weaker, but still strong enough such that low-field state remains an inhomogeneous VL. Here we track the vortex state of a moderately pinned thin films of *a*-ReZr as a function of temperature and magnetic field. We show that the inhomogeneous VL at low fields transform to an almost perfectly ordered lattice with increase in temperature or magnetic field before transforming into a VL again.

Our sample under investigation is a 20 nm thick *a*-ReZr thin film grown on oxidised silicon substrate with $T_c$ ~ 5.7 K. *a*-ReZr is a conventional extreme Type II superconductor[32] with coherence length, $\xi \sim 5.9\ nm$, and penetration depth, $\lambda \sim 800\ nm$. In this kind of samples pinning is not deliberately introduced but rather naturally originates from imperfections in the underlying substrate. Comparing the critical current densities at low fields (see Supplementary document), we observe that our film is more strongly pinned than *a*-MoGe film of the same thickness[20] where an ordered hexagonal lattice is realised at fields below 1 kOe; on the other hand, the pinning is weaker than the 5 nm thick *a*-ReZr thin film[5] where an inhomogeneous liquid state was earlier seen to encompass the entire *H-T* parameter space. The magnetic field is applied perpendicular to the film surface. Since the thickness of the film is much shorter than the characteristic bending length of vortices (typically several microns)[33] the vortex system is effectively in the 2-dimensional limit. Zero-field-cooled (ZFC) and field-cooled (FC) a.c. magnetic screening response of the sample measured down to 460 mK, show complete absence of thermomagnetic irreversibility for fields as low as 4 Oe (Fig. 1(c)), precluding the possibility of the formation a vortex glass state even at very low fields.

We image the vortex state through scanning tunnelling spectroscopy (STS). The vortex state can be imaged by measuring the tunnelling conductance (derivative of the tunnelling current with respect to voltage around a fixed dc bias voltage, $G(V) = \frac{dI}{dV}\big|_V$) while rastering the tip of a scanning tunnelling microscope (STM) on the film surface[34]. Since the core of each vortex behaves like a normal metal where the superconducting coherence peak in the local density of states (LDOS) is suppressed, the spatial map of $G(\mathbf{r}, V)$ ($\mathbf{r}$ is the position on the sample surface) shows local minima at locations where there is a vortex, when $V \approx V^{coh}$, the coherence peak voltage. This technique has been widely used to obtain large area images of the vortex solid and vortex glass states in several superconductors[11,20]. However, since STS imaging is a slow technique, imaging the VL state where vortices are moving is tricky. When the movement of vortices is faster than the acquisition time of data at each pixel, $G(V)$ at a location is time averaged over the motion of vortices. The $G(\mathbf{r}, V^{coh})$ map reflects the fraction of time a vortex is found at a given location. For a homogeneous VL with fast moving vortices $G(\mathbf{r}, V^{coh})$ will be flat and featureless. However, for a pinned VL the pinning potential and the collective interaction between vortices, create preferred locations where the vortices spend longer time than others. In $G(\mathbf{r}, V^{coh})$ map, these locations appear as local minima. Consequently, the $G(\mathbf{r}, V^{coh})$ map contains several minima of unequal depth whose value depends on the fraction of time a vortex spends there. Over longer time, some of the minima also change positions, as the vortices rearrange. The structure and dynamics of the pinned VL can be understood by analysing a series of $G(\mathbf{r}, V^{coh})$ maps as a function of time. This technique was applied to elucidate the formation of inhomogeneous VL in ultrathin amorphous superconducting films[5]. Here we apply the same technique to investigate the vortex state in the 20 nm thick *a*-ReZr thin film.

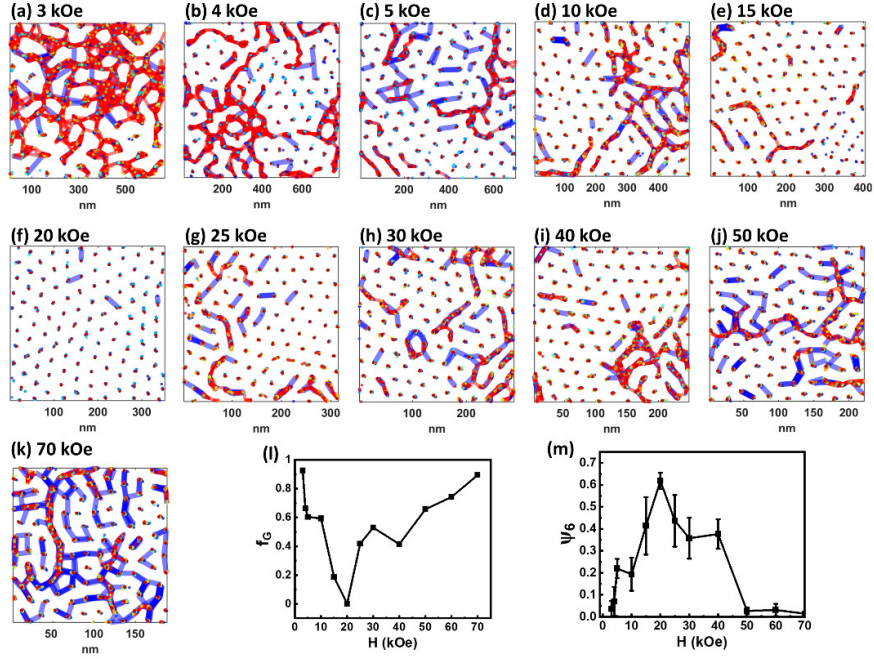

**Fig 2| Re-entrant order-disorder transformation with Magnetic field. (a)-(k)** Evolution of global motion paths of the vortices for various magnetic fields *at T* = 460 mK. The area of the maps is adjusted around a common central point to keep ~120 vortices within the field of field of view. We observe a gradual evolution from an inhomogeneous VL to a nearly ordered vortex solid as the field increased from 3-20 kOe; above this field the vortices transform again into an inhomogeneous VL. **(l)** Fraction of vortices connected to motion paths, $f_G$, as a function of magnetic field at 460 mK. **(m)** Variation of six-fold orientational parameter $\Psi_6$ with magnetic field showing re-entrant behavior with maximum at 20 kOe; error bars are obtained from standard deviation across the 20 consecutive images taken at each field.

From STS we observed an inhomogeneous VL state down to 1 kOe at 460 mK. Fig. 1(d)-(f) show 3 representative normalised conductance ($G_N(r)$) maps at coherence peak bias from a series of 20 consecutive images acquired at 3 kOe, 460 mK (For the data at 1 kOe, 460 mK, see Supplementary material). The normalisation is done such that the highest and lowest value of $G_N(r)$ across all 20 images correspond to 0 and -1 respectively. Each image is acquired over 20 min. We observe the presence of several local minima (shown with blue circles) that correspond to the position preferred by vortices. However, the value of conductance at the minima and in between two minima vary widely: Deep minima correspond to locations where vortices are strongly localized, and shallower minima correspond to points where vortices spend some time before hopping to an adjacent shallow minimum. The total number of minima is 10-15% larger than the expected number of vortices indicating that some adjacent shallow minima correspond to a single vortex hopping back and forth between those sites. On the other hand, inspecting different images we find that the positions of some minima also change with time. This happens since some vortex while hopping between two minima get trapped in a pinning potential in between, thus changing the configuration of the preferred sites in its vicinity in the next image due to intervortex interactions. This can be seen in Fig. 1(g) where the local minima from all 20 images are shown. To obtain the global motion path of vortices in this inhomogenous VL we follow a procedure similar to that described in ref 5. First, in each individual image we Delaunay triangulate the set of local minima to uniquely determine the nearest neighbours and connect any two adjacent minina for which $G_N(r) > -0.9$ at the minima, and $G_N(r) < -0.5$ for all points in-between. This accounts for the complete hops from one minimum to another. On the other hand, to account for the motion due to incomplete hops we join any two points in Fig. 1(g) that are separated by a distance less that $0.5a_H$ where $a_H = 1.075(\Phi_0/B)^{1/2}$ (where, $\Phi_0 = 2.068 \times 10^{-15}$ Wb is the flux quantum and $B$ is the magnetic field in Tesla) is the expected lattice parameter for a hexagonal lattice, but exclude any isolated segment that is shorter than $0.5a_H$ since they correspond to a meandering motion of the same vortex about its mean position. Combining all the motion paths from 20 images we get a network of global motion paths for the vortices shown in Fig. 1(h). At 3 kOe, except for a small number of isolated vortices nearly all vortices are connected to the global motion path, confirming the formation of an inhomogeneous VL.

The central result of this paper is presented in Fig. 2 and Fig. 3. Fig. 2 shows the evolution of the pinned VL with magnetic field. In Fig. 2 (a)-(k) we show the network of global motion paths at 460 mK, from 3 kOe up to 70 kOe. The motion network becomes increasingly sparce as the field is increased from 3 kOe to 20 kOe. At 20 kOe very few vortices show any movement. Furthermore, the vortex lattice appears almost perfectly ordered. Above this field the motion path again becomes progressively denser and the inhomogeneous VL state is recovered. This is also seen from the fraction of minima connected to the motion paths, $f_G$ ( Fig. 2(l) ), which is nearly the same at 3 kOe and 70 kOe but drops close to zero at 20 kOe. An interesting observation is that while at 3 kOe the red and blue connections in the motion paths largely overlap, at 70 kOe the network is dominated by blue connections which correspond to complete hops. This is because at high fields the intervortex spacing is small and therefore a vortex getting trapped between two minima (incomplete hops) is less probable. Further evidence of this re-entrant disorder-order-disorder transformation, is obtained by computing the metric of six-fold orientational order[35], $\Psi_6 = \frac{1}{N}\langle\sum_{k,l} e^{[6i(\phi_k-\phi_l)]}\rangle$, on the set of minima in each $G_N(r)$ map; here, $\phi_k$ is the angle between a fixed direction in the plane and the $k$ - th bond connecting two nearest neighbour minima. For a perfectly ordered hexagonal lattice, $\Psi_6 = 1$, whereas for an isotropic liquid or an amorphous lattice, $\Psi_6 \sim 0$. Fig. 2(m) shows the variation of $\Psi_6$, averaged over 20 images, as a function of *H*. At 3 kOe, $\Psi_6 \sim 0.035$ consistent with an inhomogeneous isotropic VL. With increase in field $\Psi_6$ increases, exhibits a sharp peak at 20 kOe where $\Psi_6 \sim 0.62$, and then decreases slowly up to 40 kOe before abruptly dropping to zero above 50 kOe. Between 25 kOe to 40 kOe we observe that the motion paths are strongly aligned along the principal axes of the underlying hexagonal lattice which is a signature of strong hexatic correlations[20,36] in the inhomogeneous VL. This hexatic correlation is destroyed above 50 kOe marking a transition to an isotropic but still inhomogeneous VL. To summarise, we clearly observe 3 states: The low field inhomogeneous VL, the high-field inhomogeneous VL and a nearly perfectly ordered vortex solid-like state in between. Similar magnetic field evolution of the vortex state at 2 K is shown in the Supplementary document.

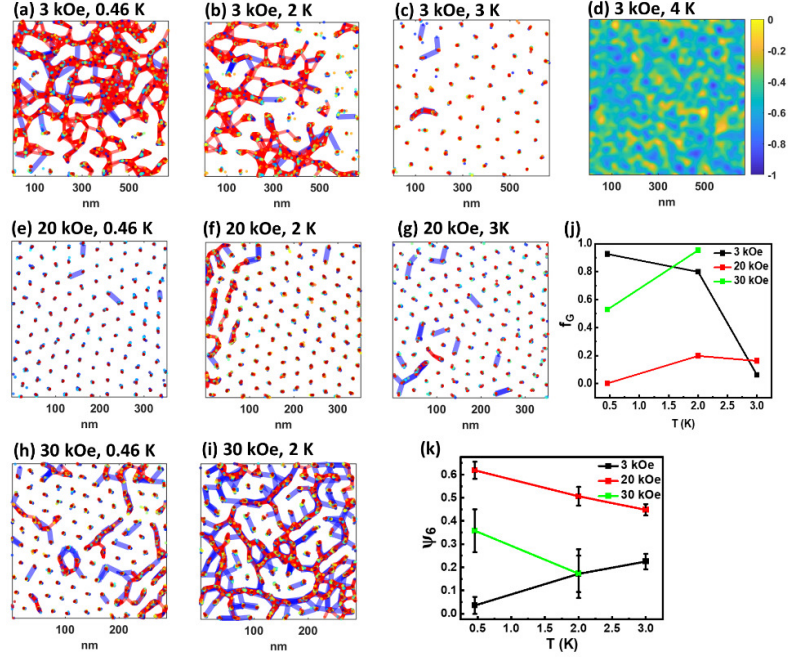

**Fig 3| Re-entrant liquid-solid-liquid transformation of the vortex state with Temperature. (a)-(c)** Global motion path at 3 kOe from 0.46 K to 3 K; the inhomogeneous VL transforms into an ordered vortex solid-like state in this temperature range. **(d)** Conductance map for 3 kOe, 4 K where the vortex lattice has completely melted; here minima are blurred due to rapid vortex motion and we cannot reliably identify individual minima anymore. Evolution of global motion path of with temperature at **(e)-(g)** 20 kOe and **(h)-(i)** 30 kOe; at high fields the vortex state progressively disorders with increase in temperature. **(j)** Variation of $f_G$ with temperature for 3 kOe**,** 20 kOe and 30 kOe respectively. **(k)** Variation of orientational order parameter $\Psi_6$ with temperature for 3 kOe**,** 20 kOe and 30 kOe respectively showing re-entrant behavior at 3 kOe and regular disordering for 30 kOe.

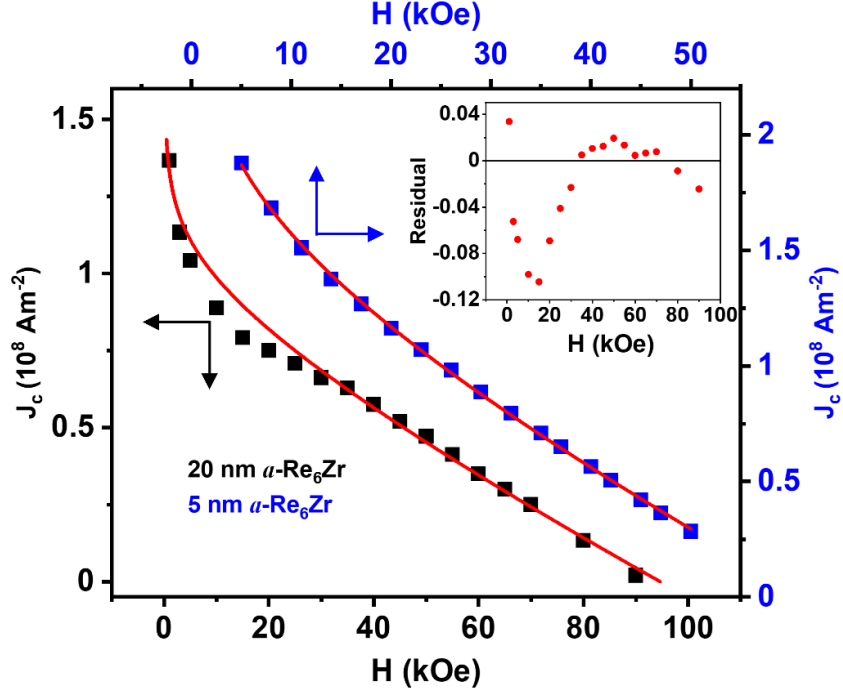

**Fig 4| Magnetic field dependence of critical current density.** Variation of $J_c$ with $H$ at 460 mK for 20 nm and 5 nm thin films of $a$-ReZr. Red lines are fits with the theoretical expression $J_c \propto H^{n-1}(1-\frac{H}{H_{c2}})$ expected for individual pinning. For 20 nm thin film, $J_c$ deviates from this variation (with n=0.91) between 5-40 kOe, where the vortex lattice gets ordered. In contrast, for the 5 nm sample (ref. 5), where the vortices form an inhomogeneous VL at all fields, this variation (with n=0.89) holds for the entire magnetic field range. The inset shows the difference between the experimental points and the fitted curve for the 20 nm sample.

    The evolution of the vortex state with temperature at fixed magnetic fields is shown in Fig. 3. When we start from the low field inhomogeneous VL at 3 kOe, the vortex motion gradually decreases, and the vortices get more ordered with increase in temperature up to 3 K. (Fig. 3(a)-(c)). This can also be seen from $\Psi_6$ which increases from 0.035 to 0.22. As the temperature is increased further to 4 K the vortex lattice melts into an isotropic VL. Here the $G_N(r)$ maps become blurred due to the rapid movement of vortices, and we can no longer identify local minima from the individual image (Fig. 3(d)). This re-entrant behaviour with temperature is not observed if one starts from the vortex solid-like state at 20 kOe (Fig. 3(e)-(g)) or the high-field inhomogeneous VL at 30 kOe (Fig. 3(h)-(i)). Here with increase in temperature the vortex state gets more disordered both in terms of increase in mobile vortices (Fig. 3(j)) and decrease in $\Psi_6$ (Fig. 3(k)).

    We now investigate signatures of these transformations on the transport properties of the superconducting film. One main distinguishing feature of a VL is the appearance of finite linear resistance well below the critical current, which is zero for a vortex solid (or vortex

glass). Unfortunately, we cannot probe this regime in our sample since voltage drops below our resolution limit. Another property that is intimately linked to the order in the vortex state is the critical current density, $J_c$, at which the Lorentz force on the vortices is equal to the pinning force, $F_p = J_c B$. For a disordered vortex state where each vortex is individually pinned $F_p$ obeys the general form[37,38], $F_p \propto h^n(1-h)$, where $n \approx 0.5 - 2$ depending on the specific pinning mechanism and $h = H/H_{c2}$. When vortices develop order, the rigidity does not allow all the vortices to be on pinning centres and $J_c$ falls below this value. Fig. 4 shows that at 460 mK, $J_c$ follows this expected magnetic field variation with $n = 0.91$ when we consider the $J_c$ values for $H < 4$ kOe and $H > 45$ kOe (for determination of $J_c$ see Supplementary material). However, between 5 – 40 kOe, where the vortex lattice order develops, $J_c$ falls below this curve. For comparison we have also plotted $J_c$ at the same temperature for the 5 nm films studied in ref. 5, where the vortices are in an inhomogeneous VL state at all fields. Here, $J_c - H$ can be well described by with the individual pinning form with $n = 0.89$ over the entire magnetic field range. Similar signatures in the magnetic field variation of the thermal activation barrier for the vortices is shown in the Supplementary material.

To summarise, using STS imaging we have conclusively demonstrated the existence of re-entrant order-disorder transformations of the VL both as function of magnetic field and temperature. However, there are some striking differences with theoretical predictions. First, the boundaries of the re-entrant transformations are not sharply defined as expected for a melting transition; rather the inhomogeneous VL gradually transforms into a near-perfectly ordered state and the solid-like state is observed at one specific field (or a very narrow range of magnetic fields). Secondly, the low field VL is expected when the intervortex spacing is much larger than $\lambda_L$. For our sample this criterion would be met for $H \lesssim 30$ Oe, whereas we observe the inhomogeneous VL up to a field that is at least 2 orders of magnitude larger. These differences can be related to the effect of pinning. The presence of pinning can broaden a sharp melting transition. In addition, since random pinning also tends to destroy the crystalline order, it shrinks the ordered crystalline phase in the *H-T* parameter space. This has been observed in 3-dimensional NbSe$_2$ samples[27,12], where with increase in pinning both the low field and high-field amorphous vortex glass phase expands in the *H-T* parameter space at the cost of the crystalline vortex solid. For the 2-dimensional vortex lattice in thin films where a vortex glass is not expected, pinning expands the inhomogeneous VL instead. This needs to be confirmed in future theoretical investigation. This also raises the natural question on whether low field VL is realised as an intrinsic feature in very weakly pinned samples. Since this state would be expected there at very low fields, it will have to be investigated using different techniques that allows imaging of the vortex state over much larger area. At the same time, it would be interesting to explore if re-entrant melting can be observed in similar two-dimensional pinned systems, such as skyrmion lattices and colloidal crystals.

**Material and Methods**

**Sample growth:** The 20 nm *a*-ReZr thin film was grown on Si/SiO$_2$ substrate through pulsed laser ablation of a Re$_6$Zr bulk target using 246 nm KrF excimer laser. The substrate was kept at room temperature during growth. Further details of sample growth and characterisation have been reported in ref. 32). To avoid surface contamination, the film was directly transferred from the deposition chamber into a ultra-high-vacuum suitcase and transferred in the low temperature STM without exposure to air. For transport measurements, the same sample was subsequently covered with a 1.5-nm-thick Ge capping layer to prevent oxidation, and then patterned into a Hall bar using argon ion beam milling.

**STS imaging of the vortex state:** The vortex state was imaged using a home-built low-temperature scanning tunnelling microscope[39] operating down to 350 mK, and fitted with a 90 kOe superconducting solenoid. We used an electrochemically etched normal metal tungsten tip. Conductance maps were obtained by measuring the tunneling conductance ($G(V) = dI/dV$) over the surface at a fixed bias voltage ($V$ ~ 1.44 mV) close to the superconducting energy gap. Each image took approximately 12 min to acquire. The precise position of the local minima in the conductance maps were determined after Fourier filtering the images to reduce noise. The nearest neighbours for each local minimum were determined by Delaunay triangulating[11] the set of points corresponding to the local minima in each image.

**Measurement of a.c. magnetic shielding response:** The a.c. magnetic shielding response is a variant of a.c. susceptibility measurement, where the superconducting film is sandwiched between a primary and a secondary coil such that the mutual inductance between the two coils depends on the degree of shielding by the film. For thin films, it is more sensitive than the conventional a.c. susceptibility measurement, where the sensitivity gets limited by the small volume of the superconductor. When the vortices form a vortex glass state, the superconductor exhibit strong history effects[11,40,41], and the FC state where vortices are all trapped in pinning potentials show a stronger shielding response than in the ZFC state. Our measurements performed with a peak a.c. excitation ~ 7 mOe at 30 kHz show the complete absence of any history effects down to 4 Oe.

## Acknowledgements

We thank Amit Ghosal, Vadim Geskenbein and Chandan Dasgupta for useful discussion. This work was financially supported by Department of Atomic Energy, Govt. of India.

## Author Contributions

RD performed the STS measurements and analysed the data. SS performed the transport measurements and analysed the data. SS and SB performed the magnetic shielding measurements. RD, SS and JJ prepared the sample. PR conceived the problem, supervised the experiments and wrote the paper with input from all authors.

[34] Suderow, H., Guillamón, I., Rodrigo, J. G., and Vieira, S., Imaging superconducting vortex cores and lattices with a scanning tunneling microscope, Supercond. Sci. Technol. 27 063001 (2014).

[35] Hattel S. A., and Wheatley, J.M., Flux-lattice melting and depinning in the weakly frustrated two-dimensional XY model, Phys. Rev. B 51, 11951 (1995).

[36] Ash, B., Chakrabarti, J., and Ghosal, A., Spatio-temporal correlations in Coulomb clusters, EPL 114, 46001 (2016).

[37] Dew-Hughes, D., Flux pinning mechanisms in type II superconductors, Phil. Mag. 30, 293 (1974).

[38] Matshushita, T, Flux Pinning in Superconductors, Springer-Verlag Berlin Heidelberg 2007.

[39] Kamlapure, A., Saraswat, G., Ganguli, S. C., Bagwe, V., Raychaudhuri, P. and Pai, S. P., A 350 mK, 9 T scanning tunneling microscope for the study of superconducting thin films on insulating substrates and single crystals, Rev. Sci. Instrum. 84, 123905 (2013).

[40] Banerjee, S. S., Patil, N. G., Ramakrishnan, S., Grover, A. K., Bhattacharya, S., Ravikumar, G., Mishra, P. K., Chandrasekhar Rao, T. V., Sahni, V. C., Higgins, M. J., Metastability and switching in the vortex state of 2H-$NbSe_2$, Appl. Phys. Lett. 74, 126 (1999).

[41] Marziali Bermúdez, M., Eskildsen, M. R., Bartkowiak, M., Nagy, G., Bekeris, V., and Pasquini, G., Dynamic Reorganization of Vortex Matter into Partially Disordered Lattices, Phys. Rev. Lett. 115, 067001 (2015).

# Supplementary Information

**Inverse melting and re-entrant transformations of the vortex lattice in amorphous Re$_6$Zr thin film**

Rishabh Duhan*, Subhamita Sengupta*, John Jesudasan, Somak Basistha and Pratap Raychaudhuri[1]
*Tata Institute of Fundamental Research, Homi Bhabha Rd, Mumbai 400005, India.*

## I. Filtering $G(r, V^{coh})$ maps

Filtering of $G(r, V^{coh})$ maps is done to remove the noise in the image. Figure 1S(a) shows a representative raw image. We first perform 2D Fast Fourier Transform (FFT) of this image (Fig. 1S(b)). The FFT shows diffused Bragg spots corresponding to the disordered vortex lattice, along with a diffuse background at higher $q$-values corresponding to high frequency noise. To remove the noise, we suppress this intensity at high $q$ values as shown in Fig. 1S(c). We then take Inverse FFT to get filtered image (Fig. 1S(d)).

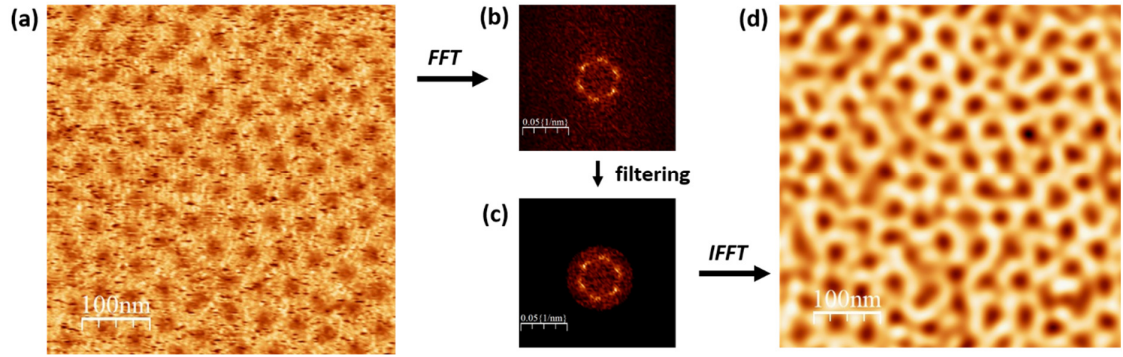

**Fig. 1S** (a) A raw image taken using STS. (b) 2D FFT of image. (c) Filtered FFT by suppressing high frequency noise. (d) Filtered image obtained by taking IFFT of (c).

## II. Pinned vortex liquid at 1 kOe, 460 mK

At 460 mK we observe the inhomogeneous VL at 1 kOe. Fig. 2S (a)-(d) shows four $G_N(r)$ maps taken consecutively over the same area. Since the vortex density is low we captured these images over a larger area, i.e. 800 nm × 800 nm, to have sufficient number of vortices within the field of view. Consequently, each of these images were captured over 24 min. We observe well resolved minima in each image. Here, since the distance between minima is large, we do not observe any complete hop of vortices from one minimum to another. However, the position of some minima changes due to incomplete hops. Fig. 2S(e) shows positions of minima obtained from all 4 images alongside with motion path connecting them.

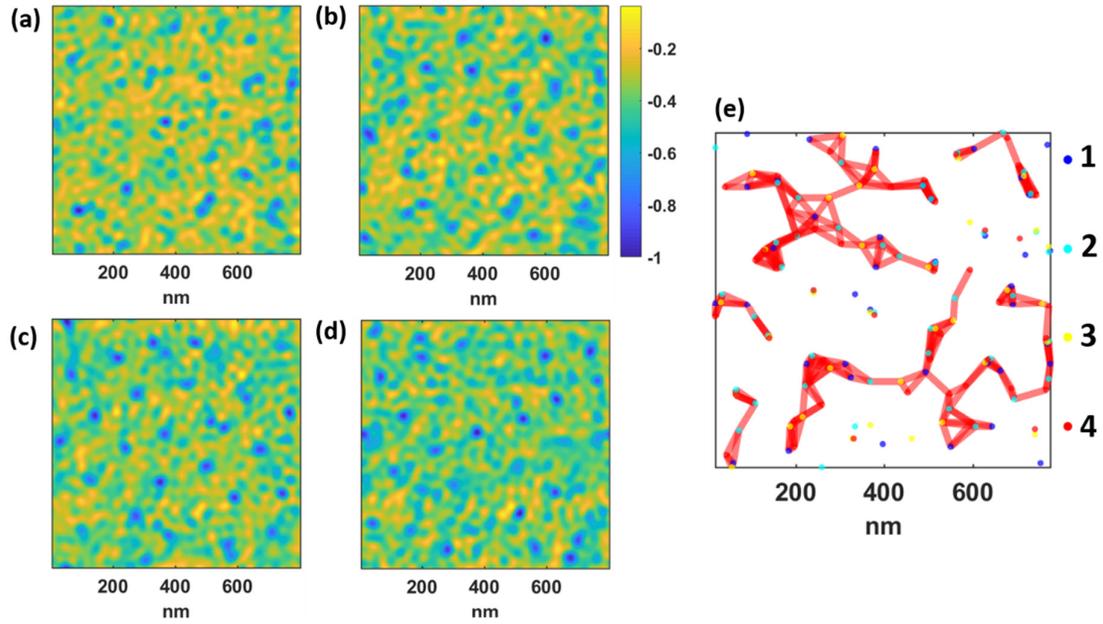

**Fig. 2S** (a)-(d) $G_N(r)$ maps taken at 1 kOe and 460 mK. (e) Position of $G_N(r)$ minima for four consecutive images along with motion path of vortices.

### III. Re-entrant behavior of VL at 2K

The re-entrant transformation of vortex state is also observed at 2 K. Fig. 3S(a) shows formation of inhomogeneous liquid at 3 kOe along with the global motion path of vortices. As we increase the field, we can see vortex state getting more ordered along with decrease in motion till 20 kOe (Fig. 3S(b)-(c)). At 30 kOe, it again forms an inhomogeneous liquid state (Fig. 3S(d)). Same can be seen in Fig. 3S(e) where $\Psi_6$ peaks and $f_G$ shows a minimum at 20 kOe.

### IV. Determination of critical current, $I_c$

Pulsed *I-V*, with a rectangular pulse of width 50 ms and interval of 6 s respectively, is taken to avoid heating at high current. Fig, 4S(a) shows pulsed I-V curves for 5 kOe, 15 kOe and 30 kOe at 460 mK. Zoomed in I-V curves for various magnetic fields at 460 mK is shown in Fig. 4S(b).

Usually, with increase in applied current, the vortex state eventually goes to Bardeen Stephen flux flow state which is characterized by flux flow resistance $R_{ff} = R_N(\frac{H}{H_{c2}})$, where $R_N$ is normal state resistance. Thus, we get a linear region in *I-V* curves from where we extrapolate to obtain $I_c$ from the intercept on current axis. However, in thin films, formation of inhomogeneous vortex state complicates the scenario, and one does not get a clear linear region in *I-V* characteristics (See ref. 5). Instead, there is a rounding off of the *I-V* curves due to vortex creep which extends till depairing current is reached (Fig. 4S(a)). Therefore, here we use a simpler criterion to determine $I_c$, namely, the current at which the voltage reaches 30 μV. Fig.

4(b) shows zoomed in I-V curves for various magnetic fields. Value of current where the horizontal dashed line at 30 µV intersects the *I-V* curves marks the criterion for determining critical current (Fig. 4S(b)).

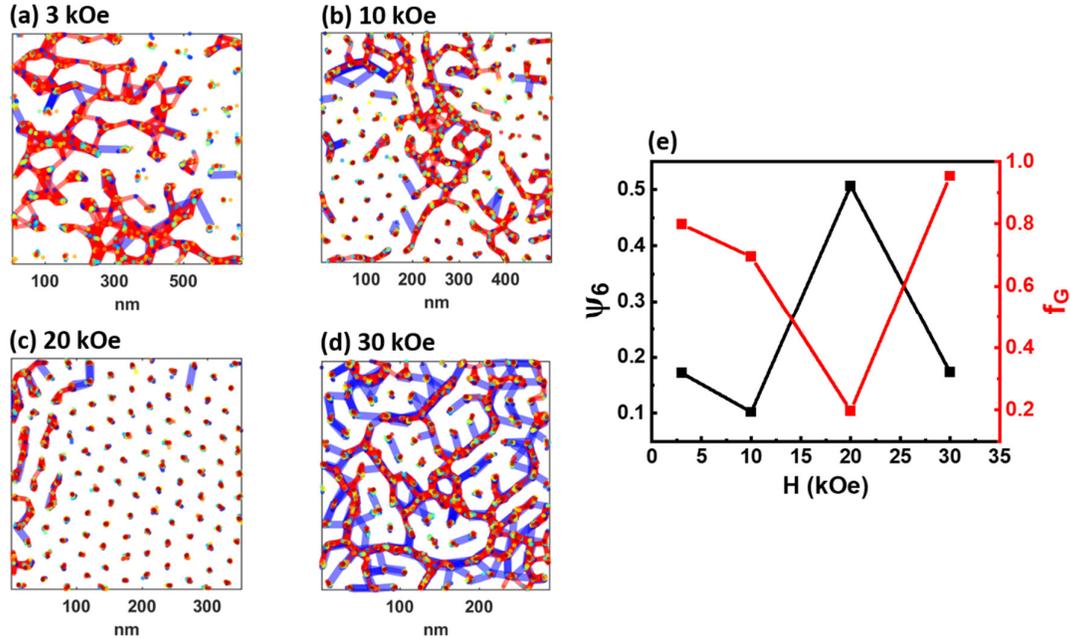

**Fig. 3S** (a)-(d) Evolution of global motion path of vortices with magnetic field at 2 K. (e) Variation of corresponding $\Psi_6$ and $f_G$ with magnetic field at 2 K.

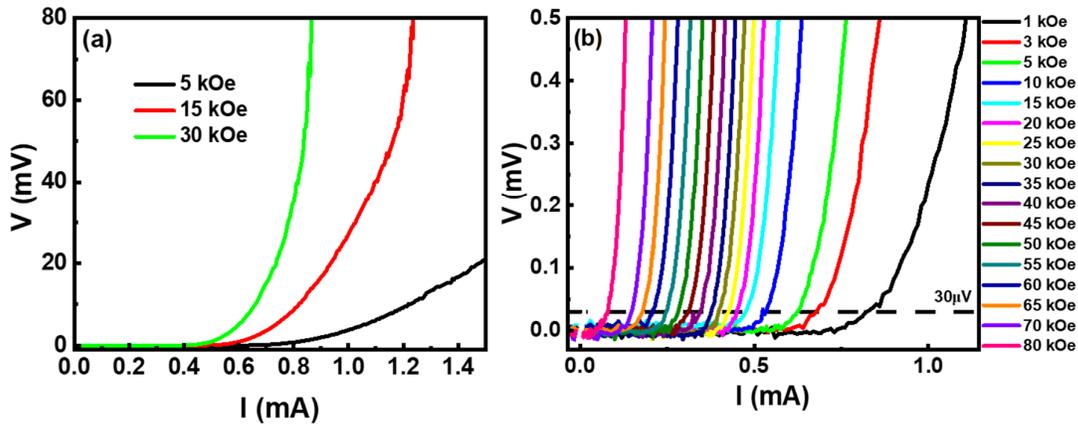

**Fig. 4S** (a) Pulsed *I-V* curves for 5 kOe, 15 kOe and 30 kOe respectively taken at 460 mK. (b) Zoomed in *I-V* curves for various magnetic fields at 460 mK. Horizontal dashed line is draw at V = 30 µV. $I_c$ at a field is determined from the value of *I* where the horizontal dashed line intersects with the respective *I-V* curve.

## V. Comparison of pinning strength

In terms of pinning the 20 nm $a$-ReZr film used in this study is intermediate between the very weakly pinned 20 nm thick $a$-MoGe film reported in ref. 20 and the strongly pinned 5 nm $a$-ReZr film used in ref. 5. To compare the pinning strength of these films we plot $J_c - H$ at 460 mK for these 3 films using the same criterion described in Section IV (Fig. 5S). Since collective effects are small at low fields, the low-field $J_c$ value is a good indicator of pinning strength. We observe that below 40 kOe, $J_c$ for the 20 nm $a$-ReZr film used here is indeed larger than the 20 nm thick $a$-MoGe film but smaller than the 5 nm $a$-ReZr film.

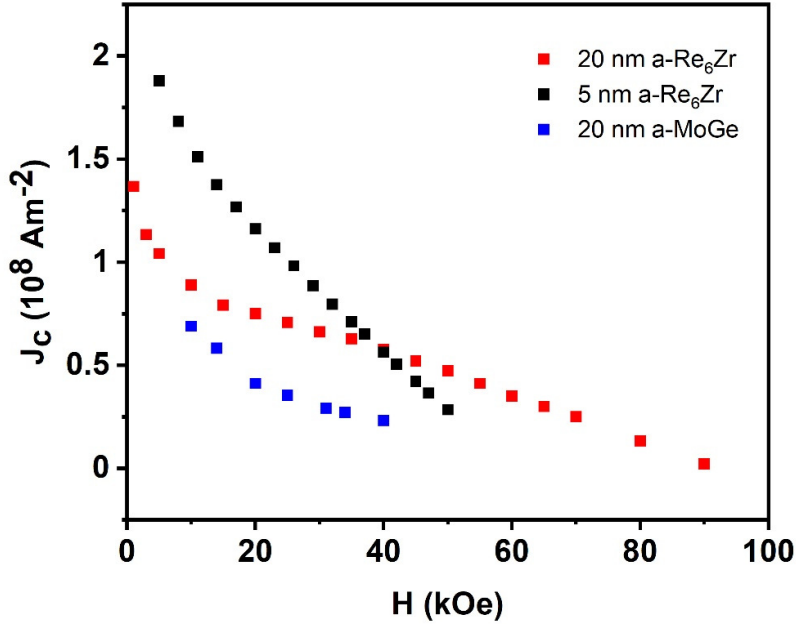

**Fig. 5S.** Variation of critical current density $J_c$ with magnetic field for 5 nm thick $a$-ReZr, 20 nm thick $a$-ReZr and 20 nm thick $a$-MoGe thin films respectively.

## VI. Signature of reentrant transition in the thermal activation barrier for vortices, $U$

In the main paper we have shown signature of the re-entrant transition on the magnetic field variation of $J_c$. Here we show that similar signature is also seen in effective pinning barrier, $U$, for thermally activated motion for vortices.

To determine $U$, in Fig. 6S(a) we plot the logarithm of the resistance resistance in ohms, $\ln(R)$, as a function of $1/T$. At low temperatures the resistance follows a simple activated behaviour, $R = R_0 e^{-U(H)/kT}$ (where $k$ is the Boltzmann constant). In Fig. 6S(b) we plot $U$, (extracted from the slope of $ln(R)$ vs $\frac{1}{T}$) as a function of $H$ in semi-log scale. We observe that for $H \leq 5\ kOe$ and $H \geq 40\ kOe$, the data fall on a straight line such that the magnetic field variation is described by a single functional form, $U(H) = U_0 ln(H_0/H)$ where $H_0 = 90.3\ kOe \sim H_{c2}$. This functional form is observed in many thin films and is expected when transport is dominated by thermal activation of dislocation pairs. On the other hand, in the magnetic field range $5 \, kOe < H < 40 \, kOe$, where the vortex lattice gradually gets ordered, $U$ is larger than what is expected from this curve. This increase results from the increase in the shear modulus ($C_{66}$) as the vortex lattice gets ordered which makes the formation of dislocation pairs more expensive (see ref. 25). In contrast for 5 nm *a*-ReZr thin film investigated in ref. 5 where vortices form an inhomogeneous liquid over the entire range of magnetic field, the functional form $U(H) = U_0 ln(H_0/H)$ captures the variation of U over the entire magnetic field range (See Fig. 1(a) of Ref. 5).

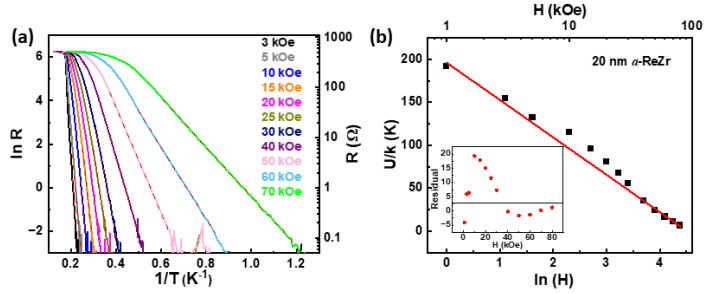

**Fig. 6S.** (a) Resistance versus temperature for the 20 nm thick *a*-ReZr film plotted as ln(R) (R in Ohms) versus 1/T; the dashed line are linear fits to these curves showing that the variation at low temperature follows a simple activated behavior. $U$ is extracted from the linear slope of the fit. The right-hand axis shows the resistance in log-scale. (b) Variation of $U$ with ln(H) (H in kOe). The red line is a fit to, $U(H) = U_0 ln(H_0/H)$, for the data in the range $H \leq 5 \, kOe$ and $H \geq 40 \, kOe$. The inset shows the difference between the data and the fit.